\documentclass[prl,reprint,twocolumn]{revtex4-1}
\usepackage{graphicx}
\usepackage{amsmath,amssymb}
\usepackage{float}
\usepackage{color}
\usepackage[normalem]{ulem}
\usepackage{siunitx}
\usepackage{layouts}
\newcommand{\Ecrit}{E_\text{crit}}

\begin{document}
\title{Realising Single-Shot Measurements of Quantum Radiation Reaction in High-Intensity Lasers}
\author{C.~D.~Baird$^1$, C.~D.~Murphy$^1$, T.~G.~Blackburn$^2$, A.~Ilderton$^3$, S.~P.~D.~Mangles$^4$, M.~Marklund$^2$, and C.~P.~Ridgers$^1$}
\affiliation{$^1$York Plasma Institute, Department of Physics, University of York, Heslington, York YO10 5DQ, UK\\$^2$Department of Physics, Chalmers University of Technology, SE-41296 Gothenburg, Sweden\\
$^3$Centre for Mathematical Sciences, University of Plymouth, PL4 7AA, UK\\
$^4$Blackett Laboratory, Imperial College London, South Kensington, London SW7 2BZ, UK}

\begin{abstract}

Collisions between high intensity laser pulses and energetic electron beams are now used to measure the transition between the classical and quantum regimes of light-matter interactions. However, the energy spectrum of laser-wakefield-accelerated electron beams can fluctuate significantly from shot to shot, making it difficult to clearly discern quantum effects in radiation reaction, for example. Here we show how this can be accomplished in only a single laser shot. A millimeter-scale pre-collision drift allows the electron beam to expand to a size larger than the laser focal spot and develop a correlation between transverse position and angular divergence. In contrast to previous studies, this means that a measurement of the beam's energy-divergence spectrum automatically distinguishes components of the beam that hit or miss the laser focal spot and therefore do and do not experience radiation reaction.

\end{abstract}

\maketitle

Bright, energetic radiation is produced across the electromagnetic spectrum when high-intensity lasers irradiate matter, due to the violent acceleration of electrons induced by the laser fields~\cite{Corde2013}.
The next generation of lasers will be sufficiently intense that recoil forces from this emission, known as radiation reaction (`RR'), will dominate the dynamics of the plasmas they create~\cite{Bell2008,Tamburini2010,Bulanov2011,Ridgers2012}.
When the energy of an individual photon of the emitted radiation becomes comparable to that of the electron, we must account for quantum effects on radiation reaction, for which there is no complete theoretical description (in the highly multiphoton regime)~\cite{Ritus1985,DiPiazza2012}.
This makes experimental validation of current models of quantum radiation reaction crucial to our understanding of the behaviour of plasmas created by next generation lasers, and to realising their many applications, which include hard X-ray sources~\cite{Kneip2010, Taphuoc2012, Chen2013, Powers2013}, compact electron accelerators~\cite{Tajima1979,Ting1997,Amiranoff1998,Malka2002,Mangles2004} and ion acceleration~ \cite{Schwoerer2006,Snavely2000,Hatchett2000,Zepf2003,McKenna2007}.

\begin{figure}[t!]
\includegraphics[width=0.45\textwidth]{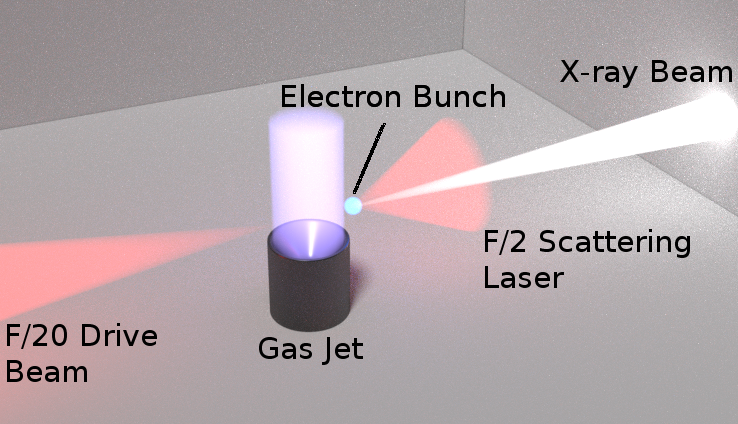}
\caption{\label{fig:ICS} A schematic of an all-optical inverse Compton scattering setup. An F/20 drive laser (intensity $\sim$\,\SI{e19}{\watt\per\square\centi\meter}) is focused into a supersonic gas jet, producing an electron bunch via wakefield acceleration. These electrons then collide with a counter-propagating F/2 laser pulse (of high intensity $\sim$\,\SI{e21}{\watt\per\square\centi\meter}). The scattered electrons produce a bright X-ray beam.}
\end{figure}

The peak intensities of current laser systems ($\sim$\,\SI{e21}{\watt\per\square\centi\meter}) are not sufficient to elicit radiation reaction (RR) effects in stationary targets. However, by pre-accelerating electrons to GeV-scale energies, for example, by laser wakefield acceleration~\cite{Leemans2006, Kneip2009,Clayton2010,Leemans2014}, RR regimes become accessible. The geometry required is similar to experiments previously used to probe Thomson and Compton scattering in the nonlinear regime~\cite{Chen1998, Chen2013, Sarri2014, Yan2017}, see Fig.~\ref{fig:ICS}. Recent experiments have shown that detection of radiation reaction is achievable on current facilities~\cite{Cole2018, Poder2017}; however due to the shot-to-shot variation of both the electron bunch and the colliding laser pulse, it was not possible to distinguish between classical and quantum (stochastic) effects on the electron motion. Here we propose a solution to this problem by incorporating a pre-interaction drift which causes the electrons' transverse momentum to become correlated to their transverse position in the bunch. After the collision, the electrons retain their initial spectral characteristics at the edges, allowing for on-shot comparison of the pre- and post-interaction spectra.

\begin{figure*}[t!]
\centering
	\includegraphics[width=0.45\textwidth]{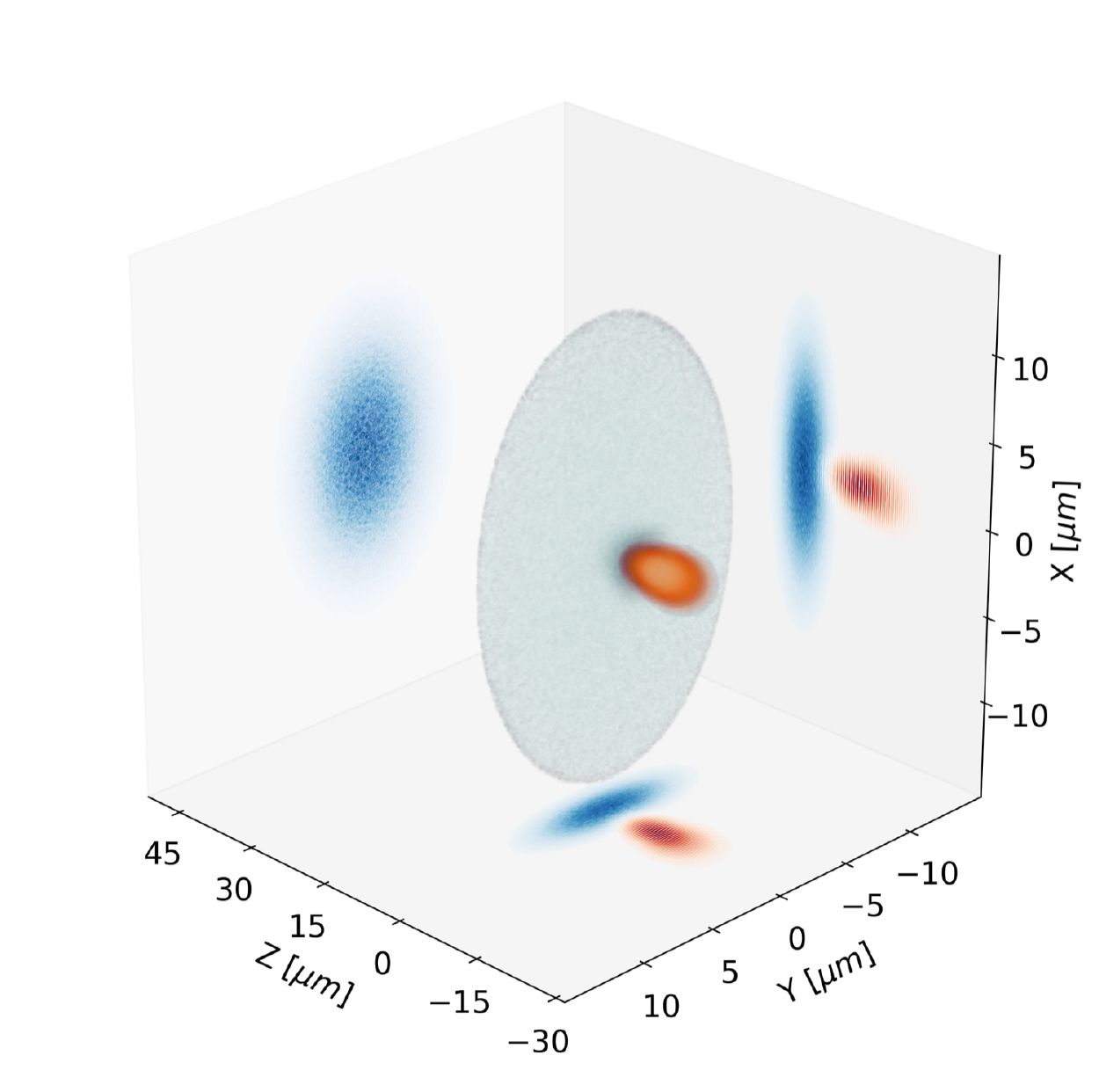}
	\includegraphics[width=0.45\textwidth]{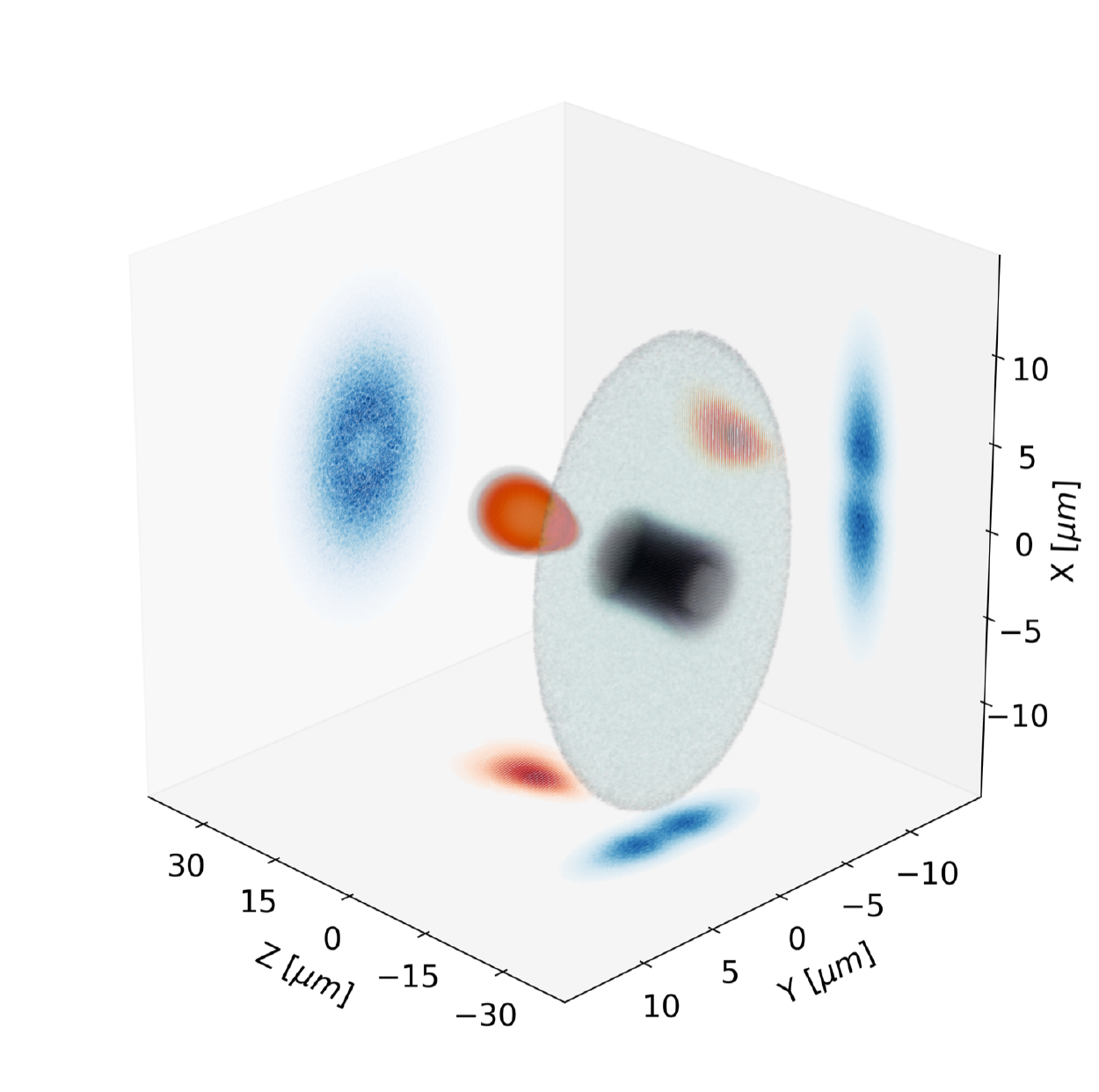}
	\caption{\label{fig:simsetup} Simulation data showing the laser pulse as it focuses through the centre of the electron bunch. The dark region indicates where electrons have lost energy due to radiation reaction. The cross-sections show the electron energy density (blue) and the laser intensity profile (red). Simulation performed using EPOCH. The spatial grid resolution was 10 cells per micron and the electron bunch was represented by $10^8$ macro-particles.}
\end{figure*}

The importance of quantum effects on radiation reaction is quantified by the parameter
\begin{equation}
\chi = \frac{\gamma\sqrt{(\vec{E}+\vec{v}\times\vec{B})^2 - (\vec{v}\cdot\vec{E})^2}}{\Ecrit} \simeq 0.1 \frac{\gamma}{1000} \frac{a_0}{20} \left(\frac{\lambda_L}{\mu\mathrm{m}}\right)^{-1}
\label{eq:eta}
\end{equation}

where the electron has velocity $\vec{v}$ and Lorentz factor $\gamma$, $\Ecrit=1.38\times10^{18}$\ Vm$^{-1}$ is the critical field of QED~\cite{Sauter1931}, and $\lambda$ and $a_0$ are the laser wavelength and strength parameter, related to the laser intensity through $I_0 = (\pi c/2)(m_e c^2 a_0 / e \lambda_L)^2$.

($\chi$ may be interpreted as the electric field in the rest frame of the radiating electron or positron.) The final expression in (\ref{eq:eta}) is valid for the specific case of the collision of an electron beam with a counter-propagating laser pulse. As $\chi$ approaches unity radiation reaction must be described in a quantum framework. For $\chi \sim 0.1$, the energy of emitted photons becomes a significant fraction of the emitting electron's energy and photon emission becomes stochastic~\cite{Blackburn2014}, rather than continuous as in the classical case. Equation (\ref{eq:eta}) demonstrates that to reach $\chi>0.1$ one needs $E>$\,\SI{500}{\mega\electronvolt} and $I>$\,\SI{e21}{\watt\per\square\centi\meter}.

Quantum corrections to the radiation spectrum, which guarantee that no photon is emitted with energy greater than the electron, reduce the power emitted compared to the classical case~\cite{Erber1966}.

The stochastic nature of the emission means that the electrons may move into classically inaccessible regions of phase space~\cite{Shen1972,Duclous2011}: in the colliding beams scenario,

quantum effects can lead to a broadening of the energy spectrum where a classical treatment can only result in  narrowing~\cite{Neitz2013,Yoffe2015,Vranic2016a,Ridgers2017}, increased emission of hard photons~\cite{Blackburn2014}, a transverse broadening of the electron bunch~\cite{Green2014} and `quenching' of emission~\cite{Harvey2017}.

See~\cite{Dinu2016} for quantum effects beyond stochasticity.

In order to measure RR effects in electron spectra, it is important that significant damping occurs during the interaction. We define `strong' RR to correspond to an electron losing 10\% of its initial energy per laser cycle.  In the interaction of an electron beam with a counter-propagating laser pulse, we can predict the required parameters from the condition $\psi:= 0.12(\gamma/1000)(a_0/10)^2>1$~\cite{Koga2005,Thomas2012}. Therefore, if $\gamma >1000$ (energy$>500$\,MeV) and $a_0>30$ we reach the regime of strong RR.

We model quantum effects, including RR, using the now-standard approach based on the `locally constant field approximation'. The basic assumption is that at high intensity the formation time of any quantum process is so short that it may be treated as an instantaneous event occurring in a field which is effectively constant. This allows quantum processes to be incorporated into classical particle-in-cell (PIC) codes as stochastic emission events. For a review see~\cite{Gonoskov2015}. This model has been implemented within the 3-dimensional PIC code EPOCH~\cite{Arber2015}, using a Monte Carlo algorithm. Details of the implementation can be found in Ref.~\cite{Ridgers2014}.

We simulated the interaction of an energetic electron bunch with a counter-propagating, high-intensity laser using EPOCH. The bunch had a central energy of \SI{1}{\giga\electronvolt}, with an RMS spread of \SI{50}{\mega\electronvolt}. It was distributed according to a 3D Gaussian number density profile with a peak of \SI{1.87e23}{\per\cubic\meter} and e-folding distances of $6\times 4 \times 4$ microns in the $x$, $y$ and $z$ directions respectively, where the laser was polarised in the $x$ direction. This elongated shape was specifically chosen to model the known spreading of laser-wakefield-generated bunches in the laser polarisation direction~\cite{Mangles2006}. The divergence profile of the bunch was taken to be a Gaussian shape, with FWHM of \SI{5}{\milli\radian}. The laser parameters were chosen to model a potential experiment on the Astra Gemini laser~\cite{Hooker2006}. The laser pulse propagated in the $z$ direction, focused to a diffraction-limited spot of width \SI{2}{\micro\meter} and had a peak focused intensity of \SI{1e21}{\watt\per\square\centi\meter}, a pulse length of \SI{44}{\femto\second} (1/e$^{2}$ in intensity) and a central wavelength of \SI{800}{\nano\meter}, which equates to an $a_0$ of 21.5.

In the simulations we varied the distance the electron bunch propagated before the interaction, which we refer to as `drift'. The purpose of doing so was to establish a correlation between the transverse position and momentum of the electrons. This drift was incorporated into the simulations by first initialising, and then redistributing the electrons by extrapolating their starting positions based on the divergence angle (neglecting space charge effects), i.e.~$x_{f} = (p_x/p_z) x_{i} d$, where $d$ is drift.  
With these initial conditions, we reach $\psi \simeq 1$, corresponding to the radiation dominated regime~\cite{Thomas2012}. Further, $\chi\simeq 0.25$ for the interaction, so quantum RR effects will be present.

\begin{figure}
\centering
\includegraphics[width=0.45\textwidth]{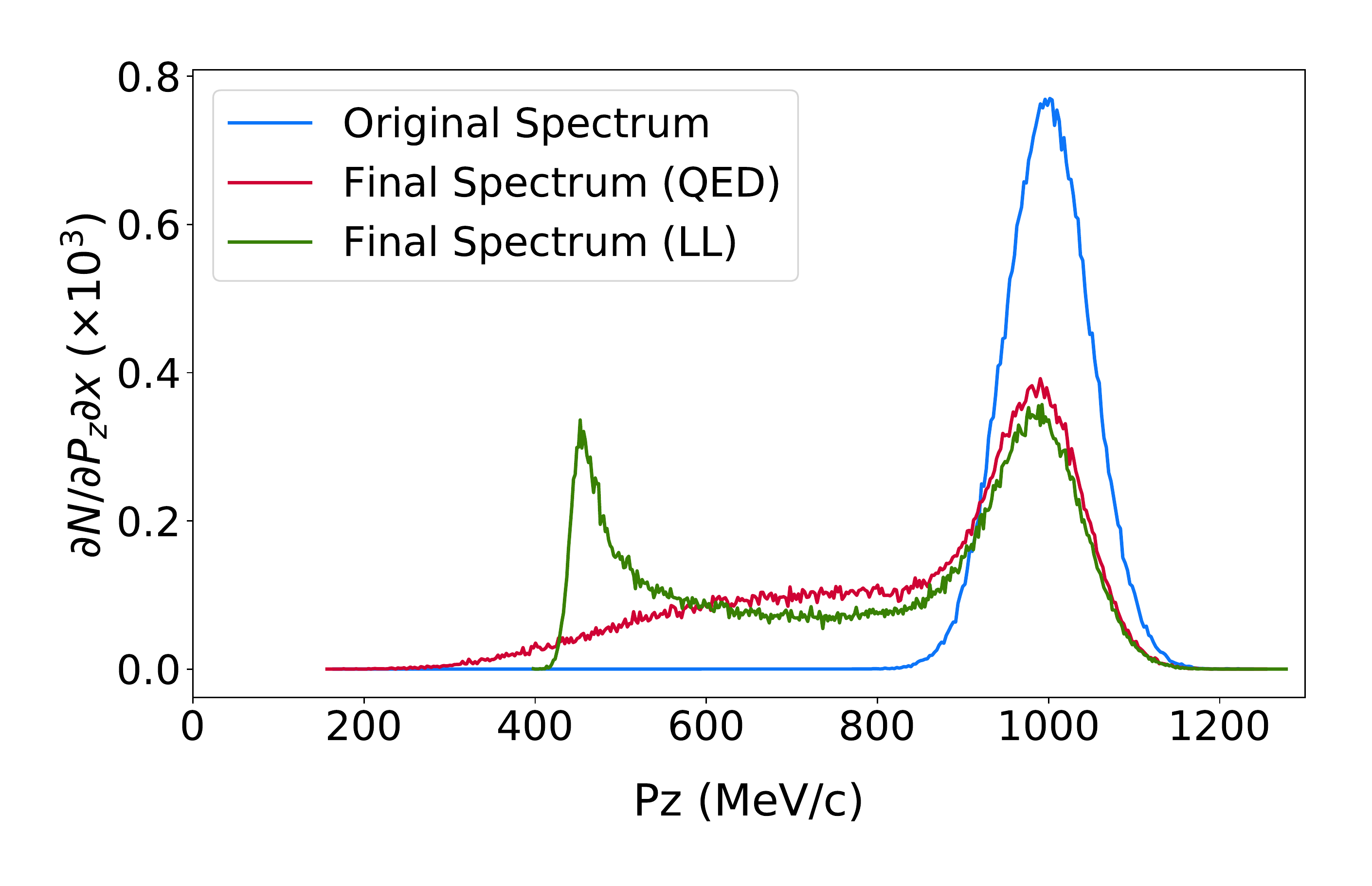}
\caption{Initial electron spectrum (blue), alongside the post-interaction spectra including QED effects (red). The emission process causes significant recoil in the electron population, resulting in a decrease in energy and an increase in spread. The classical prediction (green), using the Landau-Lifshitz model, shows bunching of the electrons at the low-energy end, as expected.}
\label{fig:post_prof}
\end{figure}
Fig.~\ref{fig:post_prof} shows the electron spectrum immediately following the interaction. Around \SI{50}{\percent} of the electrons have emitted hard photons and as a result experienced RR, lowering the peak and introducing a long, low energy tail into the spectrum. The discrepancy between the classical model (based on the Landau-Lifshitz (LL) equation~\cite{Landau1975, Burton2014}) and QED is most apparent in the low-energy region.

\begin{figure}
\centering
        \includegraphics[width=0.45\textwidth]{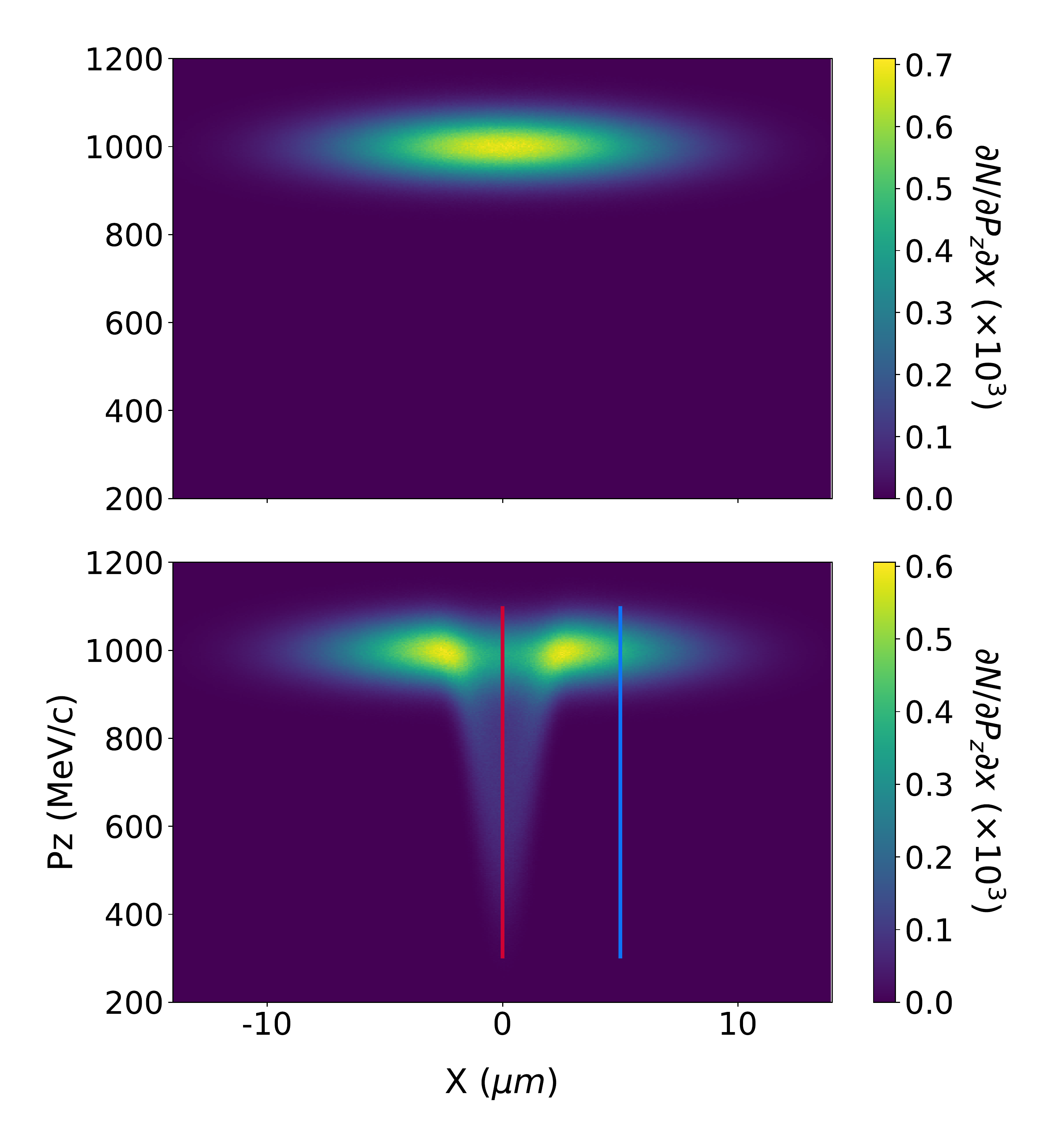}

\caption{Phase space representation of the electron bunch before (a) and after (b) interaction with the laser pulse; there is a significant shift to lower energy in the central region marked by the red line, whereas the edge regions (blue line) have not interacted. Modelled using quantum radiation reaction.}
\label{fig:synscr}
\end{figure}

Fig.~\ref{fig:synscr} shows phase-space representations of the electron bunch before, Fig.~\ref{fig:synscr}(a), and immediately after the interaction, Fig.~\ref{fig:synscr}(b). It can be seen that the central region of the electron bunch, i.e. where the bunch overlaps with the laser pulse, has experienced radiation reaction, resulting in a long tail of low energy electrons. The edges of the electron bunch, however, have remained unchanged since the width of the electron bunch is larger than that of the focused laser pulse. The fact that the central region of the image gives the electron spectrum after interaction and the edge regions retain the original electron spectrum would, crucially, allow us to determine the effect of radiation reaction on the spectrum on a shot-by-shot basis, regardless of variations in the pre-interaction spectrum.

This is already a significant result. However, in a real experiment the electron spectrum cannot be measured immediately after the interaction. Resolving the energy spectrum of the bunch requires propagation through a spectrometer magnet, the length of which may extend for tens of centimetres. Moreover, the screen must be placed some distance from the magnet to optimise energy resolution (usually a metre or more).  Over this distance, the divergence of the electron bunch causes the spectrum to blur, such that the shifted electrons spread across the full width of the bunch.

We can, however, solve this problem by varying the initial drift in the manner discussed above. Increasing the drift distance has the effect of reducing the divergence in the central region, where the interaction occurs, and thus the large propagation distance through a spectrometer causes less blurring of the spectrum. The spectrum in the central region then retains the signature of RR, whereas the edge spectrum resembles the original, as desired.

\begin{figure}
	\centering	
	\includegraphics[width=\columnwidth]{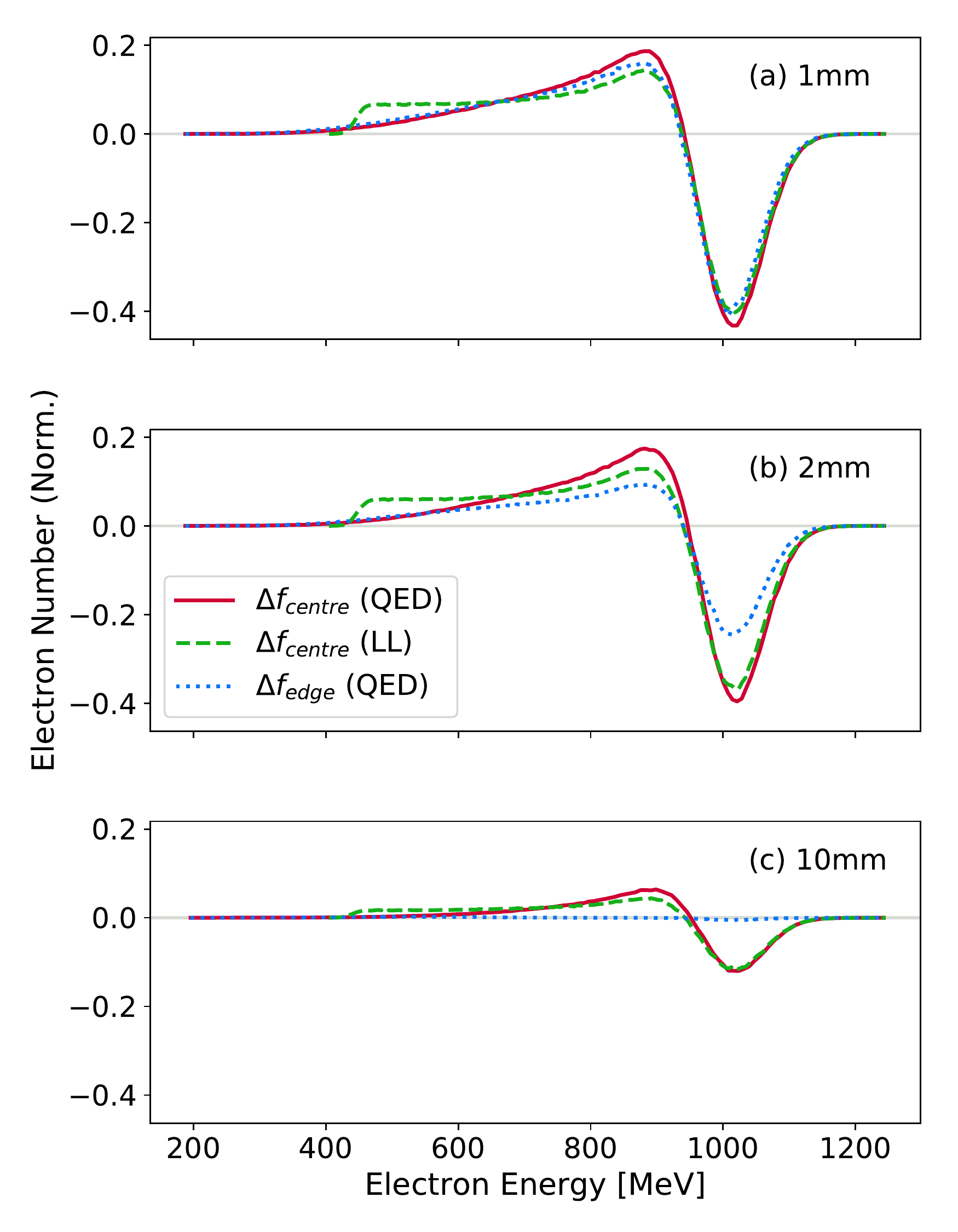}
	\caption{Electron energy spectra taken from the central region (red), and the edge region (blue). The central region from the LL interaction is also shown (green). The profiles shown are generated by subtracting the spectrum with no laser interaction (laser off), from that with laser interaction (laser on). The initial drift distance is (a) \SI{1}{\milli\meter}; (b) \SI{2}{\milli\meter}; (c) \SI{10}{\milli\meter}. It can be seen that, at $d$ = \SI{10}{\milli\meter}, the edge spectrum (blue) closely resembles that of the original. The edge spectrum from the LL interaction is not shown for clarity, but follows the same pattern as the QED interaction.}
	\label{fig:lonoff}
\end{figure}

To confirm that the edge region does indeed represent the original, we compare the post-interaction spectra at the centre and edge of the bunch, $f_{\mathrm{laser}}$, to control spectra taken from an electron bunch which has not interacted with the laser, $f_{\mathrm{no\ laser}}$. We expect that the spectrum at the edge of the screen should match the pre-interaction spectrum, allowing us to contrast with the spectrum at the centre. It can be seen in Fig.~\ref{fig:lonoff} that as the initial propagation distance, $d$, increases, the edge spectrum does indeed tend toward that of the control, i.e.~pre-interaction, spectrum (so $\Delta f:= f_{\mathrm{laser}} - f_{\mathrm{no\ laser}} \simeq 0$). Furthermore, comparison with the central region shows that the signature of the interaction is indeed retained in the spectrum, albeit reduced somewhat due to the decreased electron density as the bunch propagates. 

There are two competing effects in play here; the divergence of the electron bunch causes it to expand as it propagates, and so the fraction of electrons in the interaction region decreases as a function of distance travelled, while the correlation between (transverse) position and momentum increases with distance. This suggests that there is an optimum drift distance where the fraction of electrons interacting is sufficient for measurement, but also where the original spectrum can be deduced from the edges of the bunch. Identifying this optimal distance would maximise our ability to measure the effect of RR shot-to-shot.

\begin{figure}
    \centering
    \includegraphics[width=0.45\textwidth]{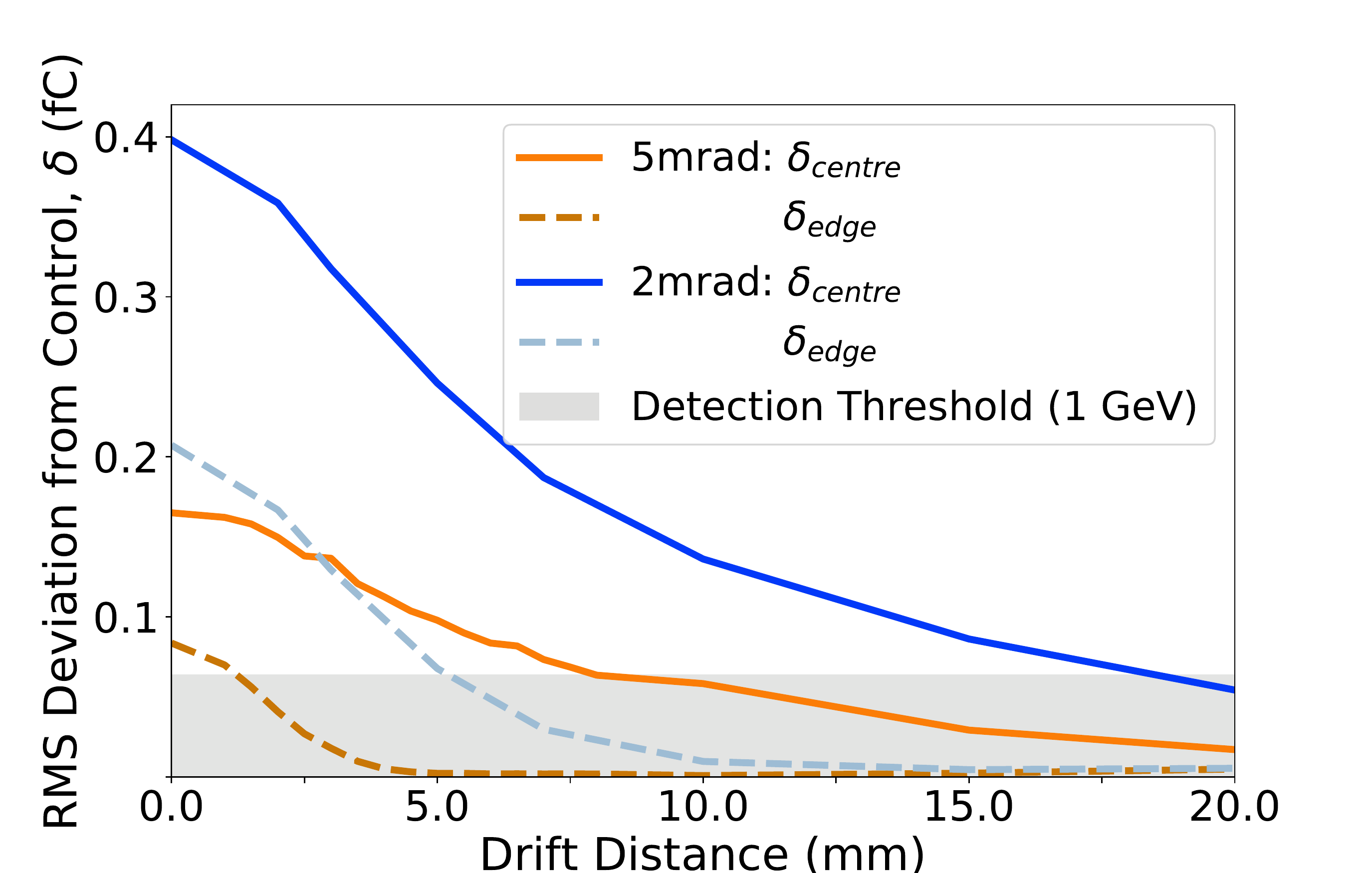}
    \caption{Plot of the values of $\delta$ in the central and edge regions as a function of drift distance for initial beam divergence of \SI{5}{\milli\radian} (orange lines) and \SI{2}{\milli\radian} (blue lines). The optimum drift distance occurs when $\delta_{edge}$ falls below the detector threshold of $\sim$ \SI{10}{\femto\coulomb\per\square\milli\meter}. The detector threshold shown is for high-sensitivity image plate as in Ref.~\cite{Buck2010}.}
    \label{fig:optimum}
    \end{figure}

To identify the optimum distance, and understand how it is affected by the initial parameters of the electron bunch, we look at how the spectrum deviates from the pre-interaction control spectrum. We take the RMS deviation-from-control, $\delta := \sqrt{\overline{(\Delta f)^2}}$, for the central (interaction) region, and also for the edge region. Fig.~\ref{fig:optimum} shows the variation of $\delta$ in the central and edge regions as a function of drift distance $d$.

As the drift distance increases, the value of $\delta$ falls both in the centre ($\delta_{centre}$) and at the edge ($\delta_{edge}$) of the screen.  When $\delta_{edge} \simeq 0$, the post-interaction spectrum at the edge is indistinguishable from the pre-interaction spectrum, and so we can compare it to the central spectrum to determine how RR affects the spectrum shot-to-shot. Considering the experimental realisation of this measurement, we can be less strict and assume that the optimum drift distance occurs where the value of $\delta_{edge}$ falls below the detection threshold of our spectrometer screen. (Similarly, the maximum drift distance is determined by the value of $\delta_{centre}$, i.e. any spectral shift ceases to be measurable below the detection threshold.) 

Studies of various types of image plate~\cite{Buck2010} have found a lower detection threshold of around \SI{10}{\femto\coulomb\per\square\milli\meter}. Using this, we can estimate the value of $\delta_{edge}$ below which the spectrum is the same as the pre-interaction spectrum to within the limits of the detector.

To quantify the detector threshold, we considered the motion of electrons through a \SI{30}{\centi\meter}, \SI{1}{T} uniform magnetic field, to a screen located \SI{70}{\centi\meter} from the exit of the magnet. This setup approximately matches the spectrometer geometry at the Astra Gemini facility. We then used the dispersion of the electrons to transform energy values on the spectrum into positional values on the screen. By translating the value of $\delta$ into an areal density of electrons, we can directly compare it to the detection threshold: this yields the grey-shaded region in Fig.~\ref{fig:optimum}. As seen in the Figure, the optimum drift distance is between \SI{1}{\milli\meter} and \SI{6}{\milli\meter} for an initial divergence of \SI{5}{\milli\radian}, and between \SI{5}{\milli\meter} and \SI{17}{\milli\meter} for a \SI{2}{\milli\radian} divergence. The RMS deviation from control is on the order of 1000 electrons, thus is likely to present difficulties for a low-sensitivity detector, such as lanex; and indeed even for a sensitive image plate if beam divergence is high. Other, more sensitive detection methods may prove invaluable in measuring this effect.

In this paper we have presented a new approach to the experimental measurement of quantum radiation reaction effects in an inverse Compton scattering arrangement. In our setup, the electron bunch is, by design, larger than the focused laser pulse. By allowing the bunch to propagate for a short distance between production and interaction, we establish a correlation between transverse position and momentum of the electrons. This preserves the transverse structure of the bunch during transport to the spectrometer screen, allowing measurement of the post-interaction spectrum in the centre of the bunch, and the pre-interaction spectrum at the edge. Although detection of the spectral shift is made challenging due to the small number of electrons involved, it should be possible with sensitive image plates, or other detectors with close to single-particle detection efficiency.

CDB, CPR \& CDM would like to acknowledge funding from EPSRC grant EP/M018156/1. TGB and MM acknowledge support from the Knut and Alice Wallenberg Foundation. MM acknowledges support from the Swedish Research Council, grants 2013-4248 and 2017-03329.

\bibliographystyle{science}

\bibliography{library}

\begin{thebibliography}{10}

\bibitem{Corde2013}
S.~Corde, et~al., {\em Reviews of Modern Physics} {\bf 85}(1) (2013).

\bibitem{Bell2008}
A.~R. Bell, J.~G. Kirk, {\em Physical Review Letters} {\bf 101}(20), 1 (2008).

\bibitem{Tamburini2010}
M.~Tamburini, F.~Pegoraro, A.~{Di Piazza}, C.~H. Keitel, A.~Macchi, {\em New
  Journal of Physics} {\bf 12} (2010).

\bibitem{Bulanov2011}
S.~V. Bulanov, et~al., {\em Nuclear Instruments and Methods in Physics
  Research, Section A: Accelerators, Spectrometers, Detectors and Associated
  Equipment} {\bf 660}(1), 31 (2011).

\bibitem{Ridgers2012}
C.~P. Ridgers, et~al., {\em Physical Review Letters} {\bf 108}(16), 1 (2012).

\bibitem{Ritus1985}
V.~I. Ritus, {\em Journal of Soviet Laser Research} {\bf 6}(5), 497 (1985).

\bibitem{DiPiazza2012}
A.~{Di Piazza}, C.~M{\"{u}}ller, K.~Z. Hatsagortsyan, C.~H. Keitel, {\em
  Reviews of Modern Physics} {\bf 84}(3), 1177 (2012).

\bibitem{Kneip2010}
S.~Kneip, et~al., {\em Nature Physics} {\bf 6}(12), 980 (2010).

\bibitem{Taphuoc2012}
K.~{Ta Phuoc}, et~al., {\em Nature Photonics} {\bf 6}(April), 1 (2012).

\bibitem{Chen2013}
S.~Chen, et~al., {\em Physical Review Letters} {\bf 155003}(April), 1 (2013).

\bibitem{Powers2013}
N.~D. Powers, et~al., {\em Nature Photonics} {\bf 8}(1), 28 (2013).

\bibitem{Tajima1979}
T.~Tajima, J.~M. Dawson, {\em Physical Review Letters} {\bf 43}(4), 267 (1979).

\bibitem{Ting1997}
A.~Ting, et~al., {\em Physics of Plasmas} {\bf 4}(5 /2), 1889 (1997).

\bibitem{Amiranoff1998}
F.~Amiranoff, et~al., {\em Physical Review Letters} {\bf 81}(5), 995 (1998).

\bibitem{Malka2002}
V.~Malka, et~al., {\em Science (New York, N.Y.)} {\bf 298}(5598), 1596 (2002).

\bibitem{Mangles2004}
S.~P.~D. Mangles, et~al., {\em Nature} {\bf 431}(7008), 535 (2004).

\bibitem{Schwoerer2006}
H.~Schwoerer, B.~Liesfeld, H.~Schlenvoigt, K.~Amthor, R.~Sauerbrey, {\em
  Physical Review Letters} {\bf 014802}(January), 1 (2006).

\bibitem{Snavely2000}
R.~A. Snavely, et~al., {\em Physical Review Letters} {\bf 85}(14), 2945 (2000).

\bibitem{Hatchett2000}
S.~P. Hatchett, et~al., {\em Physics of Plasmas} {\bf 7}(5), 2076 (2000).

\bibitem{Zepf2003}
M.~Zepf, et~al., {\em Physical Review Letters} {\bf 90}(6), 064801 (2003).

\bibitem{McKenna2007}
P.~McKenna, et~al., {\em Plasma Physics and Controlled Fusion} {\bf 49}(12 B)
  (2007).

\bibitem{Leemans2006}
W.~P. Leemans, et~al., {\em Nature Physics} {\bf 2}(10), 696 (2006).

\bibitem{Kneip2009}
S.~Kneip, et~al., {\em Physical Review Letters} {\bf 103}(035002) (2009).

\bibitem{Clayton2010}
C.~E. Clayton, et~al., {\em Physical Review Letters} {\bf 105}(10), 3 (2010).

\bibitem{Leemans2014}
W.~P. Leemans, et~al., {\em Physical Review Letters} {\bf 113}(December), 1
  (2014).

\bibitem{Chen1998}
S.~Chen, A.~Maksimchuk, D.~Umstadter, {\em Nature} {\bf 396}(December), 653
  (1998).

\bibitem{Sarri2014}
G.~Sarri, et~al., {\em Physical Review Letters} {\bf 113}(22), 1 (2014).

\bibitem{Yan2017}
W.~Yan, et~al., {\em Nature Photonics} {\bf 11}(8), 514 (2017).

\bibitem{Cole2018}
J.~M. Cole, et~al., {\em Physical Review X} {\bf 8}(1), 11020 (2018).

\bibitem{Poder2017}
K.~Poder, et~al., {\em arXiv} {\bf 1709}(01861v2), 1 (2017).

\bibitem{Sauter1931}
Z.~Sauter, {\em Phys.} {\bf 69}, 742 (1931).

\bibitem{Blackburn2014}
T.~G. Blackburn, C.~P. Ridgers, J.~G. Kirk, A.~R. Bell, {\em Physical Review
  Letters} {\bf 112}(1), 1 (2014).

\bibitem{Erber1966}
T.~Erber, {\em Reviews of Modern Physics} {\bf 38}(4), 626 (1966).

\bibitem{Shen1972}
C.~S. Shen, D.~White, {\em Physical Review Letters} {\bf 28}(7), 455 (1972).

\bibitem{Duclous2011}
R.~Duclous, J.~G. Kirk, A.~R. Bell, {\em Plasma Physics and Controlled Fusion}
  {\bf 53}(1), 015009 (2011).

\bibitem{Neitz2013}
N.~Neitz, A.~{Di Piazza}, {\em Physical Review Letters} {\bf 111}(5), 1 (2013).

\bibitem{Yoffe2015}
S.~R. Yoffe, Y.~Kravets, A.~Noble, D.~A. Jaroszynski, {\em New Journal of
  Physics} {\bf 17} (2015).

\bibitem{Vranic2016a}
M.~Vranic, T.~Grismayer, R.~A. Fonseca, L.~O. Silva, {\em New J. Phys.} {\bf
  18}(073035) (2016).

\bibitem{Ridgers2017}
C.~P. Ridgers, et~al., {\em arXiv:1708.04511} pp. 1--14 (2017).

\bibitem{Green2014}
D.~G. Green, C.~N. Harvey, {\em Physical Review Letters} {\bf 112}(16), 1
  (2014).

\bibitem{Harvey2017}
C.~N. Harvey, A.~Gonoskov, A.~Ilderton, M.~Marklund, {\em Physical Review
  Letters} {\bf 118}(10), 1 (2017).

\bibitem{Dinu2016}
V.~Dinu, C.~Harvey, A.~Ilderton, M.~Marklund, G.~Torgrimsson, {\em Phys. Rev.
  Lett.} {\bf 116}(044801), 1 (2016).

\bibitem{Koga2005}
J.~Koga, T.~Z. Esirkepov, S.~V. Bulanov, {\em Physics of Plasmas} {\bf
  12}(093106) (2005).

\bibitem{Thomas2012}
A.~G.~R. Thomas, C.~P. Ridgers, S.~S. Bulanov, B.~J. Griffin, S.~P.~D. Mangles,
  {\em Physical Review X} {\bf 2}(041004), 1 (2012).

\bibitem{Gonoskov2015}
A.~Gonoskov, et~al., {\em Physical Review E} {\bf 92}(023305), 1 (2015).

\bibitem{Arber2015}
T.~D. Arber, et~al., {\em Plasma Physics and Controlled Fusion} {\bf 57}(11),
  113001 (2015).

\bibitem{Ridgers2014}
C.~P. Ridgers, et~al., {\em Journal of Computational Physics} {\bf 260}, 273
  (2014).

\bibitem{Mangles2006}
S.~P.~D. Mangles, et~al., {\em Physical Review Letters} {\bf 96}(21), 1 (2006).

\bibitem{Hooker2006}
C.~J. Hooker, et~al., {\em Journal de Physique IV (Proceedings)} {\bf 133}, 673
  (2006).

\bibitem{Landau1975}
L.~D. Landau, E.~M. Lifshitz, {\em {The Classical Theory of Fields}}. Elsevier,
  Oxford (1975).

\bibitem{Burton2014}
D.~A. Burton, A.~Noble, {\em Contemporary Physics} {\bf 55}(2), 110 (2014).

\bibitem{Buck2010}
A.~Buck, et~al., {\em Review of Scientific Instruments} {\bf 81}(3) (2010).

\end{thebibliography}

\end{document}